# Audio Input Generates Continuous Frames to Synthesize Facial Video Using Generative Adiversarial Networks

Zhang H.H.D.
School of Information Technology and Electrical Engineering
The University of Queensland, Qld., 4072, Australia

**Abstract**

*This paper presents a simple method for speech videos generation based on audio: given a piece of audio, we can generate a video of the target face speaking this audio. We propose Generative Adversarial Networks (GAN) with cut speech audio input as condition and use Convolutional Gate Recurrent Unit (GRU) in generator and discriminator. Our model is trained by exploiting the short audio and the frames in this duration. For training, we cut the audio and extract the face in the corresponding frames. We designed a simple encoder and compare the generated frames using GAN with and without GRU. We use GRU for temporally coherent frames and the results show that short audio can produce relatively realistic output results.*

## 1 Introduce

Human voice and facial features are obvious parts of the identity of person. These two characteristics are highly correlated in speech videos. Previous research [1-3] has established that we can use a speech signal to generate facial image. However, how to the generate a sequence of video frames aligned with the conditioning speech is still a challenge work.

In this paper, we propose a new methodology for face frames generation, using aligned audio. Building on previous work, A. Duarte et al. [1] have presented a method for generating a facial image give a speech signal. Unlike their works, however, we focus on generating a sequence of face images using relatively shorter audio signals.

For this purpose, we propose a conditional generative adversarial model (shown in Figure 1) that contains GRU in both generator and temporal discriminator. Since there is no data set that meets our requirements, for learning this self-supervised model, we reference the ideas of A. Duarte et al. [1] that get dataset from youtube and make our own datasets.

The contributions of our work are summarized as follows:

We propose a simple method to generate face video by training conditional GAN that could generate face sequence frames directly from short speech signal.

We use simple designed audio encoder and With GRU in the GAN, we compared different methods to solve temporal smooth.

For our tasks, we build a dataset containing paired signals and frames.

## 2 Related work

Generative Adversarial Networks. GANs [4], since it was proposed, have been widely used in various fields of deep learning. Image generation is one of the earliest applications of GAN. Early methods focused on generating random images [4-8]. A. Radford, et al. [6] introduce deep convolutional generative adversarial networks (DCGANs), considering the limitation of convolutional networks (CNNs) for unsupervised learning. After this, T. Salimans, et al. [7] proposed a variety of methods to improve the training process of GAN. I. Gulrajani and M. Arjovsky, et al. [5, 8]eventually came up with new algorithms that could stabilize training and get rid of problems such as mode collapse.

In previous studies on Conditional Generative Adversarial Nets (CGAN) [9], which is able to control to output of the model, there are many different conditions to generate ideal images that look like real images. Image translation is based on image to generate high resolution false image [10, 11].

Face image generation. Face generation is an important direction of image generation because face is an important feature of human recognition and is widely used. Gauthier [12] used external information as condition to generate face images achieves similar result as vanilla generative model. Antipov, et al. [13] and Wang, et al. [14] used GAN to do face aging which aims to generate fake faces with given age.

Video generation. The video generation is based on the image generation, but because each frame of the video is continuous with the characteristics of time sequence, the simple use of independently generated images will affect the temporal smooth. Chen [15] and Chan, et al. [16] respectively proposed the method of dance video synthesis. Chan proposed a discriminator for timing detection in which multiple frames are input. Chen uses music to generate dance videos, and GRU is adopted in music encoding.

There are many video synthesis papers have designed special modules for the continuity and fluidity of time [17-20].The Deep video-based performance cloning system Aberman, et al. [17] was approached with two branches, one learning the appearance and another improving the temporally coherent performance. In Dynamics Transfer GAN [18], which is considered as 2 parts, the spatial and the dynamics, the dynamic part obtains the temporal consistency, and the spatial part uses a pre-trained RNN to suppress dynamic features. The Recycle-GAN for Unsupervised video retargeting [19] adds temporal information to constraints and combines the cycle loss, recurrent loss and adversarial loss. The previous Moco-Gan [20] which guides the Recycle-GAN designed 4 subnets includes a video discriminator used for temporal smooth test and training.

Recurrent Neural Network. (RNN) It is a neural network for processing sequential data. Considering that every frame in the video is a time-correlated data, it is feasible to adopt the RNN model. However, RNN and its other forms, such as long-term Memory [21] and GRU, have few direct applications in computer vision. CNN, LSTM and GRU are combined to predict regional rainfall data and propose a new model called Trajectory GRU (TrajGRU) [22]. In this paper, we used ConvGRU and used this module in the generator and discriminator to generate continuous frames and to distinguish time smoothness.

## 3 Method

### A. Dataset build:

In order to train the model, we need the audio and the corresponding frame, however, the current data set does not meet such requirements. In [1], there is already a method to extract a video and an intermediate frame. On this basis, we further designed a method to process the video to establish my own data set. The video was captured from YouTube.

The number of frames of the video is 30 frames per second. In order to group the three consecutive frames together with the 0.1s audio, I formed a group of data. The audio format is the converted 16kHz and the Haar Featurebased Classifier is used to extract the face. In order to make people's facial changes have continuity, we excluded frames without human faces and corresponding audio, and finally for a video there is

### B. Audio encoder:

Pascual, et al. [23] and Hershey, et al. [24] designed encoders for audio. Among them, Viggish[24] is a pre-trained model that can directly extract audio features without training. However, these two methods are not suitable for our model, because these two models need to input long audio fragments, and there is no way to input 0.1 audio into 533 data. Therefore, we designed a relatively simple model, taking the normalized audio as input and transforming it into a 128-dimensional vector through several 1D convolutional layers.

### C. ConvGRU

In this paper the main form of ConvGRU was as follows in [22]:

$$Z_t = \sigma(W_{xi} * X_t + W_{hz} * H_{t-1})$$
$$R_t = \sigma(W_{xr} * X_t + W_{hr} * H_{t-1})$$
$$H'_t = f(W_{xh} * X_t + R_t \circ (W_{hh} * H_{t-1}))$$
$$H_t = (1-Z_t) \circ H'_t + Z_t \circ H_{t-1}$$

'$*$' is convolution, '$\circ$' means Handamard product and $f$ is activation. $H_t$, $R_t$, $Z_t$, $H'_t \in \mathbb{R}^{Ch \times H \times W}$ are information from the last hidden layer, reset gate, update gate and output to next GRU cell. $X_t \in \mathbb{R}^{Ci \times H \times W}$ is the input of GRU cell. $Ci$, $Ch$ are chanal sizes. Reset gate control how much the information is remained from last state and update gate control how much the the information sends to next state.

In our project, the ConvGRU cell is used in the last layers of Generator. $H \times W$ is the size of ouput frames size.

### D. GAN

The based GAN model we use is the conditional GAN. The GAN objective

$$\min_G \max_D V(D,G) = \mathbb{E}_{x \sim P_{data}(x)}[\log D(x|y)] + \mathbb{E}_{z \sim P_z(z)}[\log(1-D(G(z|y)))]$$

$y$ is the condition and $x$ is input, which is the true image. With the condition, the generator $G$ will generate image $G(z|y)$, in which $z$ is noise.

In fact, the typical model will produce individual frames of temporal artifacts. To solve this, our discriminator adopted 3 sequenced frames as input and the GAN objected is updated

$$L = \mathbb{E}_{(x,y)}[\log D(x_t, x_{t+1}, x_{t+2}, y)] + \mathbb{E}_x[1 - \log D(G(y), y)]$$

GAN contains the discriminator and generator, and structure of the ours is shown in Fig 1. The ConvGRU in discriminator with sequenced frames as inputs and finally ouputs $H_t$ to next layer while that in generator, inorder to generate frames, the input is only from last layers and $H_t$ s are frames. In both Conv block and ConvTranspose block, they contain batchnormalized layers and activation, LeakyReLU and ReLU.

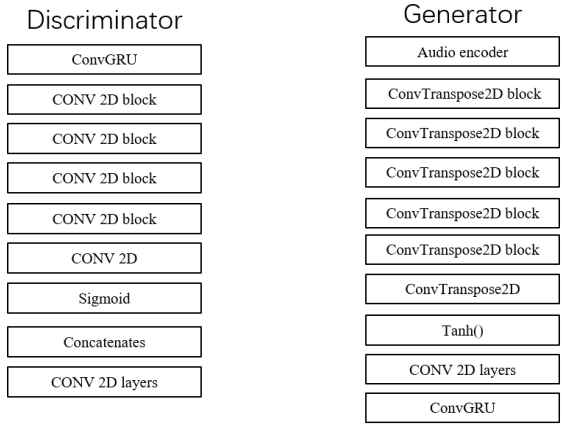

Fig 2. This is the architecture of the discriminator and generator.

## 4 Experiments and results

We compare our performance to baseline methods (wav2pix) and the model with out CovGRU.

To set up the experiments, we select one video from youtube.

Baselined model we chose is Wav2pix. Since this model cannot input too short audio, we follow the processing method of the original paper. We cut the audio into 1s and take the frame in the middle of a second and end up with 242/242 audio, frame data.

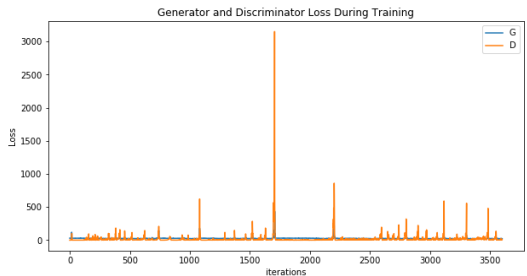

Fig 3.loss of D and G in the Baseline model training

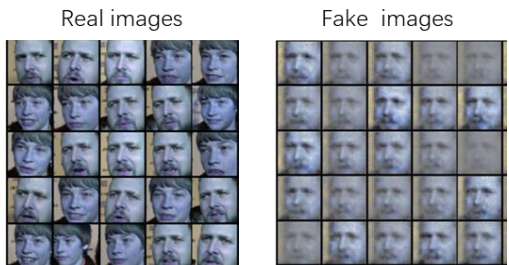

Fig 4. result of the baseline model

Ablation conditions: ConvGRU. we ablate the ConvGRU layer and to generate outputs with 9 chanals which will be separated to 3 images with 3 chanals.

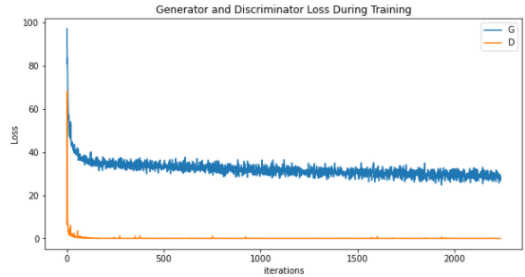

Fig 5. loss of D and G in the Baseline model training

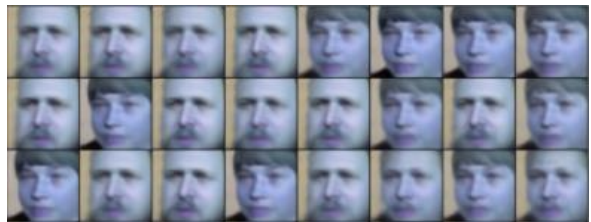

Fig 6. result of the no-GRU model

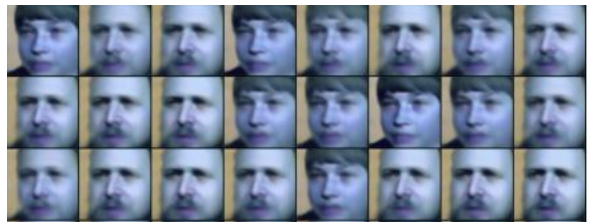

Fig 7. result of the our model

Evaluation metrics: we use three metrics to measure the quality of each frame:1）Mean Squared Error (MSE), 2) Structural Similarity (SSIM) and Learned Perceptual Image Patch Similarity (LPIPS)

Evaluation: In the *Fig 4*, *Fig 6* and *Fig 7*, there are the results of different model. Compare with the baseline, our models It seems a little bit clearer, and I think there are a couple of reasons for that. First, the training data of the basic model is insufficient. Although we train in the same video, the input audio time of the basic model is relatively long. The amount of data is obviously smaller than that of our model, and no data enhancement method is used. This problem can be seen in the change of loss during the training. To get result, we increased the number of iterations, so that the results could still see fuzzy but recognisable faces.

When doing ablation experiment, we chose basically the same training process and parameters. Our implementation is based on the PyTorch library and trained on a NVIDIA Tesla K80 GPU with 12GB memory. We kept the the hyperparameters as suggested in. the learning rate is both 0.0001 in D and G.

| Metric | Baseline | No-GRU | With-GRU |
|--------|----------|--------|----------|
| MSE    | 0.0965   | 0.1074 | 0.1034   |
| SSIM   | 0.1564   | 0.9525 | 0.9543   |
| LPIPS  | 0.5885   | 0.4778 | 0.4931   |

*Table 1. Metric comparison of different models. For SSIM higher is better. For MSE and LPIPS lower is better.*

The result shows in the Table 1 report that the baseline model have the best MSE but the SSIM and LPIPS is not good. Our model better SSIM and LPIPS. GRU has some improvement on MSE and SSIM, although it is not particularly obvious on the numerical value. This can be seen in the Fig 6 and Fig 7. The images generated by model with GRU looks real and clear.

## 5 Conclusion

In this work we introduce a method to generate speech face video with given speech and propose a simple model with ConvGRU to generate sequenced frames using short Audio as input. We took the video from YouTube and processed it, extracting three consecutive frames for every 0.1s of audio. Our experiments show that our model can generate more accurate and recognisable faces under the same video compared to Wav2pix's model. And the inclusion of ConvGRU would have made the model generation clearer.

The following research is to generate more continuous frames for longer speech, and for the identity of face, an identity recognition module can be added. The future can increase the range of generation, not limited to people's faces.

## Acknowledgment

I would like to express my deep gratitude to Professor Huang, my research supervisors, for her patient guidance, enthusiastic encouragement and useful critiques of this research work. I would also like to thank my tutor Ruihong Qiu and Zhi Chen, for their advice and assistance in keeping my progress on schedule.

Finally, I wish to thank my parents for their support and encouragement throughout my study.

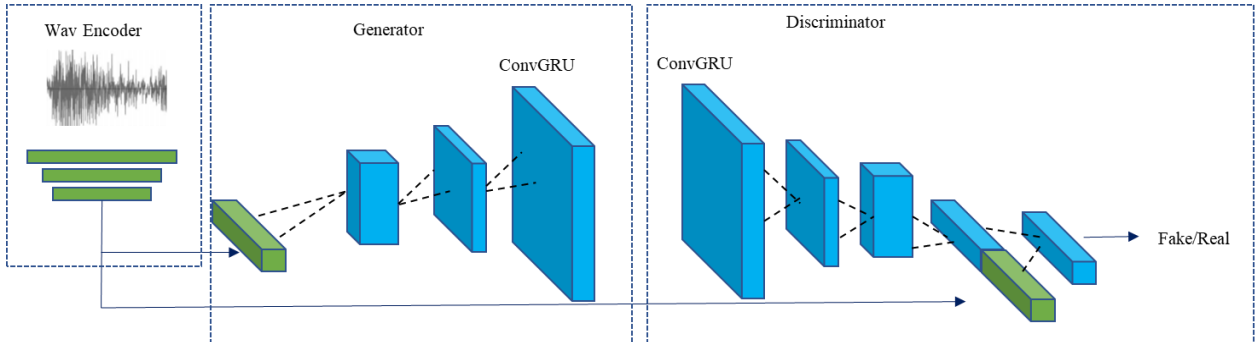

*Fig 8. The overall structure of our model.*